\documentclass[prb,a4paper,twocolumn,showpacs,superscriptaddress]{revtex4}

\usepackage{amsmath}
\usepackage{amssymb}
\usepackage{graphicx}

\newcommand{\bA}{\mathbf{A}}
\newcommand{\bB}{\mathbf{B}}
\newcommand{\be}{\mathbf{\hat{e}}}
\newcommand{\bj}{\mathbf{j}}
\newcommand{\br}{\mathbf{r}}

\newcommand{\p}{\partial}
\newcommand{\ph}{\varphi}
\renewcommand{\Im}{\mathrm{Im}}

\begin{document}

\author{E. Anisimovas}
\affiliation{Departement Natuurkunde,
Universiteit Antwerpen (Campus Drie Eiken), Universiteitsplein 1,
B-2610 Antwerpen, Belgium}
\affiliation{Semiconductor Physics Institute, Go\v{s}tauto 11, LT-01108
Vilnius, Lithuania}
\author{A. Matulis}
\affiliation{Semiconductor Physics Institute, Go\v{s}tauto 11, LT-01108
Vilnius, Lithuania}
\affiliation{Departement Natuurkunde,
Universiteit Antwerpen (Campus Drie Eiken), Universiteitsplein 1,
B-2610 Antwerpen, Belgium}
\author{F. M. Peeters}
\email{Francois.Peeters@ua.ac.be}
\affiliation{Departement Natuurkunde,
Universiteit Antwerpen (Campus Drie Eiken), Universiteitsplein 1,
B-2610 Antwerpen, Belgium}

\title{Currents in a many-particle parabolic quantum dot
under a strong magnetic field}

\date{31 August 2004}

\begin{abstract}
Currents in a few-electron parabolic quantum dot placed into a
perpendicular magnetic field are considered. We show that
traditional ways of investigating the Wigner crystallization
by studying the charge density correlation function can be
supplemented by the examination of the density-current correlator.
However, care must be exercised when constructing the correct
projection of the multi-dimensional wave function space. The
interplay between the magnetic field and Euler-liquid-like
behavior of the electron liquid gives rise to persistent and
local currents in quantum dots. We demonstrate these phenomena
by collating a quasi-classical theory valid in high magnetic
fields and an exact numerical solution of the many-body problem.
\end{abstract}


\maketitle

\section{Introduction}

The problem of the electronic structure\cite{maks00,kouw01,reimann02}
of few-electron quantum dots\cite{jacak98} still remains in the
center of attention of solid state research. Although recently
substantial achievements have been made by relying on the exact
numerical solution\cite{maarten03,yang02,mikh01} of the complicated
quantum-mechanical problem,
simplified approaches\cite{vas02} and
simple analytical models are also of
great interest and even become increasingly more
popular.\cite{lent91,tan99,mat94,yann02,yann03,anis04} Besides their
relative
simplicity, these models are attractive due to the provided
physical insight and transparent visualization of the relevant
phenomena.

The successful development of the approximation based on the
picture of a rotating electron molecule\cite{yann02,yann03}
encouraged us to look at the rotation in quantum dots more
closely. Namely, we investigated the distribution of currents
in the simplest quantum dot --- the rotating electron ring
formed in a few-electron system in a parabolic confinement and
a perpendicular magnetic field.

Bedanov and Peeters\cite{bedanov94}
predicted the crystallization of a system of classical
two-dimensional (2D) particles in an external parabolic
confinement into a
set of concentric rings. In quantum dots containing up to five
electrons only one ring is formed. An accurate quantum-mechanical
solution based e.~g.\ on the exact numerical diagonalization of the
many-electron Hamiltonian provides a circularly symmetric
distribution of the charge density. The formation of the Wigner
crystal can be seen in the density-density correlation
function\cite{maksym96,reimann00} obtained by considering the
conditional probability to find an electron at a point $\br$ given
that there is an electron at another point $\br_0$.

Recently, the analysis of various electron structures was further
stimulated by employing a description based on mean-field approaches,
such as the density-functional theory.\cite{reimann02} In this method,
the electron-electron correlation is taken into account as an
effective single-particle potential, and the correlation function
is not easily accessible. Thus, the rotational symmetry of the dot
has to be broken in some artificial way, and the electron
crystallization at high magnetic fields is made visible directly
in the electron density.
Consideration of the currents\cite{reimann99} also portrays the
presence of a crystal --- currents are circulating around the
charge density lumps.

In the present paper, we study the local currents appearing in the
vicinity of the electron lumps in a rotating Wigner molecule from
the point of view of the density-current correlation function. This
function describes the distribution of conditional currents in the
system given that there is an electron at a certain point and indeed
shows the electron crystallization. However, a certain care must be
exercised when constructing 2D projections of the
multi-dimensional space of many-body currents.

The magnetic field tries to rotate the electron system as a rigid
body, i.~e.\ with a constant vorticity. As the Schr\"{o}dinger
equation has no dissipation the ``quantum electron liquid'' is akin
to the Euler liquid in hydrodynamics and tries to rotate itself
with a fixed angular momentum, that is, with vorticity equal to zero
everywhere except at the origin. This contradiction leads to two
interesting phenomena: local currents and persistent currents. Both
of them are caused and controlled by the electron-electron
interaction. We demonstrate these phenomena by constructing a
simple quasi-classical theory employing a rotating frame in the
limiting case of high magnetic fields, and illustrate the obtained
results by comparing them to the exact numerical solution of the
considered problem.

The outline of the paper is as follows. Section II describes the
model used in our calculations. The next two Sections are
devoted to the analytical solution: in Sect.\ III we derive the
Hamiltonian in the rotating frame, and Sect.\ IV describes its
quantization. The results concerning the persistent and local
currents are presented in Sects.\ V and VI, respectively. We
conclude with a summarizing Section VII.

\section{Model}

We consider a 2D parabolic quantum dot containing
$N = 2-5$ electrons placed into a perpendicular magnetic field
$\bB$. This system is known to form a single electron ring. The
magnetic field is described in terms of its symmetric-gauge
vector potential $\bA = \frac{1}{2}[\bB \times \br]$, then using
the standard procedure\cite{maarten03} of switching to
dimensionless variables we obtain the following Hamiltonian
\begin{equation}\label{ham0}
  H = \frac{1}{2}\sum_{n=1}^N\left\{\left(-i\nabla_n
  + \frac{1}{2}[{\bf B}\times{\bf r}_n]\right)^2 + r_n^2 \right\}
  + \sum_{n,m=1\atop n>m}^N\frac{\lambda}{|{\bf r}_n - {\bf r}_n|}.
\end{equation}
Here, the energy is measured in units $\hbar\omega_0$ with
$\omega_0$ being the characteristic frequency of the confinement
potential. The coordinates are measured in units
$l_0=\sqrt{\hbar/m^*\omega_0}$ (the oscillator lengths), and the
magnetic field in units $\Phi_0/\pi l_0^2$ ($\Phi_0=\pi\hbar c/e$).
The dimensionless Coulomb coupling constant  $\lambda = l_0/a_B^*$
is expressed as the ratio of the parabolic confinement length $l_0$
to the effective Bohr radius $a_B^*=\epsilon\hbar^2/m^*e^2$.
One way to find the eigenstates of the above Hamiltonian is to perform
the ``exact'' numerical diagonalizations in the truncated Hilbert
space of all possible electron configurations as described in
e.~g.\ Ref.\ \onlinecite{maarten03}. Such calculations are feasible
for few-electron quantum dots. Concentrating on the orbital effects
we neglect the Zeeman energy but do take the degeneracy due to the
spin states into account when constructing the basis of many-body
states. This approach is supplemented by an approximate analytical
scheme valid at high magnetic fields and employing a quasi-classical
expansion in powers of $B^{-1}$ whose development is the subject of
the following Sections.

\section{Rotating frame}

Since we are looking for the ground state wave function, the
strong magnetic field term in the parentheses of Eq.~(\ref{ham0})
must be compensated by a large angular momentum $M$ that appears
in the first term of the same parentheses when the gradient
operator acts on the rotating electron ring wave function. The
easiest way to take this cancelation of terms into account is to
use the Eckardt frame\cite{eckardt} which is the main instrument
employed in the analysis of the rotation-vibration spectra of
molecules and was recently adapted to quantum
dots.\cite{maksym96}

The idea of the rotating Eckardt frame is to decouple the two
degrees of freedom related to the center-of-mass motion as well as
one more degree of freedom of the rotational motion of the system
as a whole around its center of mass. The angular velocity of the
rotating frame is chosen so that the system in the Eckardt frame
has zero total angular momentum. As we are going to apply the
expansion in $B^{-1}$ powers and are interested only in the
lowest-order harmonic approximation for the vibrations of the
Wigner crystal, the above technique can be simplified using a
frame rotating about the center of the quantum dot rather than
the center of mass.

In order to derive the Hamiltonian written in the rotating frame
we follow the suggestion of Ref.\ \onlinecite{maksym96} and start
from the classical Lagrangian
\begin{subequations}\label{lagrangian}
\begin{eqnarray}
\label{lagrangian1}
  L &=& L_{\mathrm{mag}} - V, \\
\label{lagrangian2}
  L_{\mathrm{mag}} &=& \frac{1}{2}\sum_{n=1}^N\left\{\dot{\br}_n^2
  - [\bB\times\br_n]\dot{\br}_n\right\}, \\
\label{lagrangian3}
  V &=& \frac{1}{2}\sum_{n=1}^N\br_n^2 +
  \sum_{n,m=1\atop n>m}^{N}\frac{\lambda}{|{\bf r}_n - {\bf r}_n|},
\end{eqnarray}
\end{subequations}
where $L_{\mathrm{mag}}$ is the sum of free-electron Lagrangians
in a magnetic field, and $V$ stands for the confinement and
interaction potentials.

\begin{figure}
\includegraphics[width=50mm]{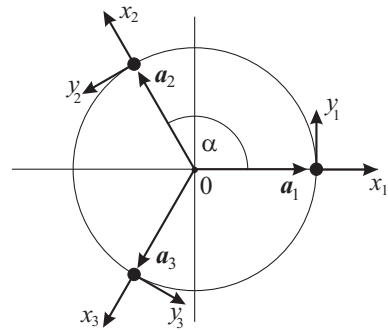}
\caption{\label{fig:loc} Local coordinates for three electrons on
a ring.}
\end{figure}

In order to transform the above expressions into the rotating
frame we follow Ref.\ \onlinecite{anis98} and introduce the local
coordinates (as shown in Fig.\ \ref{fig:loc} for the case of three
electrons).
We denote the classical equilibrium radius of the ring by $a$. The
azimuthal angles $\alpha_n = \alpha(n-1)$ with $\alpha=2\pi/N$ and
$n=1,\cdots,N$ indicate the equidistant locations of electrons on
the ring. We choose the local
axes $x_n$ to be parallel to the equilibrium location vectors
${\bf a}_n$ of the respective electrons, and $y_n$-axes are
directed along the ring in the positive (counter-clockwise)
direction. We assume that the whole frame shown in Fig.\ \ref{fig:loc}
is rotating in the positive direction with a constant angular
velocity $\dot{\chi}$. Thus, the position of the vector ${\bf a}_n$
is given by the angle
\begin{equation}\label{aangle}
  \alpha_n(t) = \alpha_n + \chi(t), \quad \chi(t) = \dot{\chi}t.
\end{equation}
We represent the electron coordinates $\br=\{x,y\}$ by a complex
number $z=x+iy$, and introduce the above-described local coordinates
by means of the transformation
\begin{equation}\label{transf}
  z_n \to \exp(i[\alpha_n+\chi])(a+z_n).
\end{equation}
This enables us to present the magnetic part of the Lagrangian
(\ref{lagrangian2}) as
\begin{equation}\label{maglg}
  L_{\mathrm{mag}} = \frac{1}{2}\sum_{n=1}^N\left\{\dot{z}_n^2
  + \left(\omega^2-B^2/4\right)|a+z_n|^2
  + 2\omega\Im(\dot{z}_nz_n^*)\right\},
\end{equation}
with $\omega = \dot{\chi} - B/2$.
This function is constrained by the condition
\begin{equation}\label{cond}
  \Im\sum_{n=1}^N(a+z_n)\dot{z}_n^* \approx
  a\,\Im\sum_{n=1}^N\dot{z}_n^*=0,
\end{equation}
which expresses the equality of the total angular momentum of
the electron system to zero. Integrating the above constraint
we rewrite it as
\begin{equation}\label{cond1}
  \sum_{n=1}^N y_n = 0,
\end{equation}
which can serve as the definition of the rotating frame.

In order to obtain the corresponding Hamiltonian we introduce the
normal modes taking into account the fact that the electron ring
is invariant with respect to rotations by a multiple of $\alpha$.
Thus, the above normal modes are just the Fourier transforms
$w_k = u_k + iv_k$
\begin{equation}\label{ft}
  z_n = \frac{1}{\sqrt{N}}\sum_{k=0}^{N-1}e^{i\alpha k(n-1)}w_k, \;
  w_k = \frac{1}{\sqrt{N}}\sum_{n=1}^{N}e^{-i\alpha k(n-1)}z_n.
\end{equation}
Note that the index $k$ numbering the normal modes runs from
$0$ to $N-1$. Inserting these expressions into Eq.~(\ref{maglg})
we obtain the following final Lagrangian
\begin{eqnarray}\label{lgmod}
 L_{\mathrm{mag}} = \frac{1}{2}\left\{\dot{u}_0^2
  + (\omega^2 - B^2/4)(\sqrt{N}a+u_0)^2\right\} \phantom{mmmm} \nonumber \\
 + \frac{1}{2}\sum_{k=1}^{N-1}\left\{|\dot{w}_k|^2
  +(\omega^2-B^2/4)|w_k|^2 + 2\omega\,\Im(\dot{w}_kw_k^*)\right\}.
\end{eqnarray}
The main advantage of this expression is that the condition (\ref{cond1})
is taken into account automatically by excluding the $v_0$ mode
which is now replaced by the rotation angle $\chi$. Next, we introduce
the canonical momenta
\begin{subequations}\label{momenta}
\begin{eqnarray}
\label{momenta1}
  M &=& \frac{\p}{\p\dot{\chi}}L_{\mathrm{mag}}
  = \frac{\p}{\p\omega} L_{\mathrm{mag}} \nonumber \\
  &=& \omega {\cal I} + \sum_{k=1}^{N-1}\Im(\dot{w}_kw_k^*), \\
\label{momenta2}
  U_0 &=& \frac{\p}{\p\dot{u}_0}L_{\mathrm{mag}} = \dot{u}_0, \\
\label{momenta3}
  U_k &=& \frac{\p}{\p\dot{u}_k}L_{\mathrm{mag}} = \dot{u}_k - \omega v_k, \\
\label{momenta4}
  V_k &=& \frac{\p}{\p\dot{v}_k}L_{\mathrm{mag}} = \dot{v}_k + \omega u_k,
\end{eqnarray}
\end{subequations}
where
\begin{equation}\label{inertia}
  {\cal I} = {\cal I}_0 + \sum_{k=1}^{N-1}|w_k|^2, \quad
  {\cal I}_0 = (\sqrt{N}a+u_0)^2
\end{equation}
is the moment of inertia of the electron ring. In order to
obtain the standard form of the magnetic part of the Hamiltonian
\begin{equation}\label{hammod}
  H_{\mathrm{mag}} = M\dot{\chi} + U_0\dot{u}_0
  + \sum_{k=1}^{N-1}(U_k\dot{u}_k+V_k\dot{v}_k) - L_{\mathrm{mag}}
\end{equation}
we have to solve Eqs.~(\ref{momenta}) for velocities. However,
aiming to arrive at the Hamiltonian in the harmonic approximation
we are entitled to make some approximations in the solution.
Namely, in Eqs.~(\ref{momenta3}) and (\ref{momenta4})
we replace the frequency by its approximate value
$\omega\approx M/{\cal I}_0$,
solve them for velocities $\dot{u}_k$ and $\dot{v}_k$, and
inserting the obtained values into Eq.~(\ref{momenta1}) we obtain
the following approximate expression for the angular velocity
\begin{equation}\label{angvel}
  \dot{\chi} \approx \frac{B}{2} + \frac{M}{\cal I} -
  \frac{1}{{\cal I}_0}\sum_{k=1}^{N-1} (V_ku_k-U_kv_k) +
  \frac{M}{{\cal I}_0^2}\sum_{k=1}^{N-1}|w_k|^2.
\end{equation}
Inserting these expressions of velocities together with the
Lagrangian (\ref{lgmod}) into Eq.~(\ref{hammod}) we arrive at the
final expression for the magnetic part of the classical Hamiltonian
\begin{eqnarray}\label{hamf}
&&  H_{\mathrm{mag}} = \frac{1}{2{\cal I}}
  \left(M+\frac{B{\cal I}}{2}\right)^2 +
  \frac{1}{2}U_0^2 \nonumber \\
&& +\frac{1}{2}\sum_{k=1}^{N-1}\left\{\left(U_k+\frac{Mv_k}
  {{\cal I}_0}\right)^2
  +\left(V_k-\frac{Mu_k}{{\cal I}_0}\right)^2\right\}.
\end{eqnarray}
The total Hamiltonian is obtained by adding the potential
(\ref{lagrangian3}) consisting of two terms that are small
in the $B\to\infty$ limit.

\section{Quantization}

The quantization of the ring rotation is trivial because the
rotation angle $\chi$ is a cyclic variable. Thus, the orbital
momentum $M$ is a constant of motion, and quantizing we simply
replace this momentum by an integer eigenvalue of the
corresponding angular momentum operator. The rotational part
of the wave function is given by
$\Psi_{\mathrm{rot}} = \exp(iM\chi)$.

Before proceeding to the quantization of the remaining vibrational modes
we have to fix the radius of the ring $a$. It is convenient to define
this parameter by minimizing the potential energy which consists
of the first term of the magnetic Hamiltonian (\ref{hamf}) and
the weak potential (\ref{lagrangian3}). Introducing the deviation
$\Delta = {\cal I} - {\cal I}_0$ and replacing the electron coordinates
in the second term of the weak potential by their equilibrium values,
we present the potential as
\begin{equation}\label{potappr}
  V = V_0 - \frac{1}{2{\cal I}_0^2}\left(M^2-\frac{B^2{\cal I}_0^2}{4}\right)
  \Delta + \frac{M^2}{2{\cal I}_0^3}\Delta^2,
\end{equation}
where the equilibrium potential is
\begin{equation}\label{potzero}
  V_0 = \frac{1}{2{\cal I}_0}\left(M+\frac{B{\cal I}_0}{2}\right)^2
  + N\left(\frac{1}{2}a^2 + \frac{\lambda}{2a}f_N\right)
\end{equation}
and the factor $f_N=\frac{1}{2}\sum_{n=1}^{N-1}|\sin(\alpha n/2)|^{-1}$
is the Coulomb energy per electron in a ring of unit radius.
Minimization of the equilibrium potential (\ref{potzero})
defines the equilibrium radius of the electron ring. As the first
term in Eq.~(\ref{potzero}) is large and the second one is small,
it is easier to obtain the estimate of the radius in two steps.
First, we equate the derivative of the first term of the potential
(\ref{potzero}) to zero and find
\begin{equation}\label{mombrel}
  -M = |M| = \frac{1}{2}B{\cal I}_0 = \frac{1}{2}NBa_0^2.
\end{equation}
This expression gives us a relation between the magnetic field
strength and the angular momentum in the ground state, and
confirms the already mentioned fact that the angular momentum
of the system grows in absolute value with increasing magnetic
field. Then, the radius of the ring $a_0$ itself is obtained by
minimizing the second (small) contribution to the potential
(\ref{potzero}):
\begin{equation}\label{weakmin}
  \frac{d}{da}\left(\frac{1}{2}a^2 + \frac{\lambda}{2a}f_N\right)
  = a - \frac{\lambda}{2a^2}f_N = 0,
\end{equation}
with the result
\begin{equation}\label{radius0}
  a_0 = (\lambda f_N/2)^{1/3}.
\end{equation}

Now the construction of the Hamiltonian in the harmonic
approximation can be completed. Taking into account that the
fulfillment of the condition (\ref{mombrel}) zeroes the second
term of the expansion (\ref{potappr}), and using the approximation
$\Delta \approx 2a_0\sqrt{N\,u_0}$ in the third term of
Eq.~(\ref{potappr}), we obtain the vibrational part of the
Hamiltonian:
\begin{equation}\label{hamvib}
\begin{split}
  H_{\mathrm{vib}} =& \frac{1}{2}(U_0^2+B^2u_0^2) \\
  +& \frac{1}{2}\sum_{k=1}^{N-1}\left\{\left(U_k+\frac{Bv_k}{2}\right)^2
  +\left(V_k-\frac{Bu_k}{2}\right)^2\right\}
\end{split}
\end{equation}
The first term of this Hamiltonian (corresponding to the breathing
mode) is just the Hamiltonian of a harmonic oscillator, and its
ground-state eigenfunction is $\exp(-Bu_0^2/2)$. The other part of
the Hamiltonian represents a collection of noninteracting two-component
modes whose Hamiltonian resembles the Hamiltonian of a 2D electron in
a perpendicular magnetic field. Thus, the corresponding ground-state
eigenfunction is $\exp(-B\sum_{k=1}^{N-1}|w_k|^2/4)$.

Collecting all parts of the wave function together and performing
the inverse Fourier transformation we obtain the electron ring
ground state wave function
\begin{subequations}\label{wff}
\begin{eqnarray}
\label{wff1}
  \Psi &=& e^{iM\chi}e^{-BK/4}, \\
\label{wff2}
  K &=& \sum_{n=1}^{N}(x_n^2+y_n^2)
   + \frac{\left(\sum_{n=1}^Nx_n\right)^2}{N}
  - \frac{\left(\sum_{n=1}^Ny_n\right)^2}{N}. \nonumber \\
\end{eqnarray}
\end{subequations}

\section{Persistent currents}

\begin{figure}
\includegraphics[width=80mm]{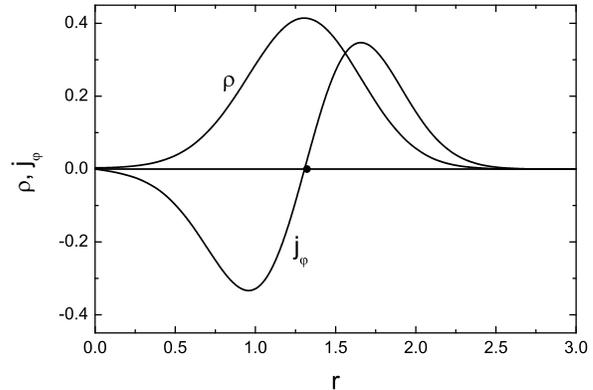}
\caption{\label{fig:dens} The radial distribution of the density
($\rho$) and azimuthal current density ($j_{\ph}$) for three
electrons in a quantum dot with $\lambda=4$ for angular
momentum $M = -18$. The formation of a ring structure is apparent.
Note that the currents flow in opposite directions on the inner and
outer edges of the ring. The peak of $\rho$ and zero of $j_{\ph}$
are close to the classical electron ring radius $a_0 \approx 1.322$
marked by the dot.}
\end{figure}

Let us proceed to the examination of the current flow in quantum
dots. In Fig.\ \ref{fig:dens}, the distributions of the charge and
current density obtained from the exact diagonalizations are plotted.
Since these
distributions are circularly symmetric, only the radial dependences
need to be shown, and moreover, the current possesses only the
azimuthal component $j_{\ph}$. Fig.\ \ref{fig:dens} is obtained for
a three-electron quantum dot with the interaction constant
$\lambda = 4$.
The angular momentum is $M = -18$ and the dimensionless magnetic
field is set to $B = 6.78$, i.~e.\ the midpoint of the magnetic
field range where the considered state is the ground state.

We see that at these parameter values the electron ring is
already well defined; the electron density at the origin drops
almost to zero and peaks close to the classical value of the
radius $a_0 = (4/\sqrt{3})^{1/3} \approx 1.322$. It is
interesting to observe that the currents flow in the opposite
direction on the inner and the outer circumference of the ring
and vanish on the classical radius.

\begin{figure}
\includegraphics[width=80mm]{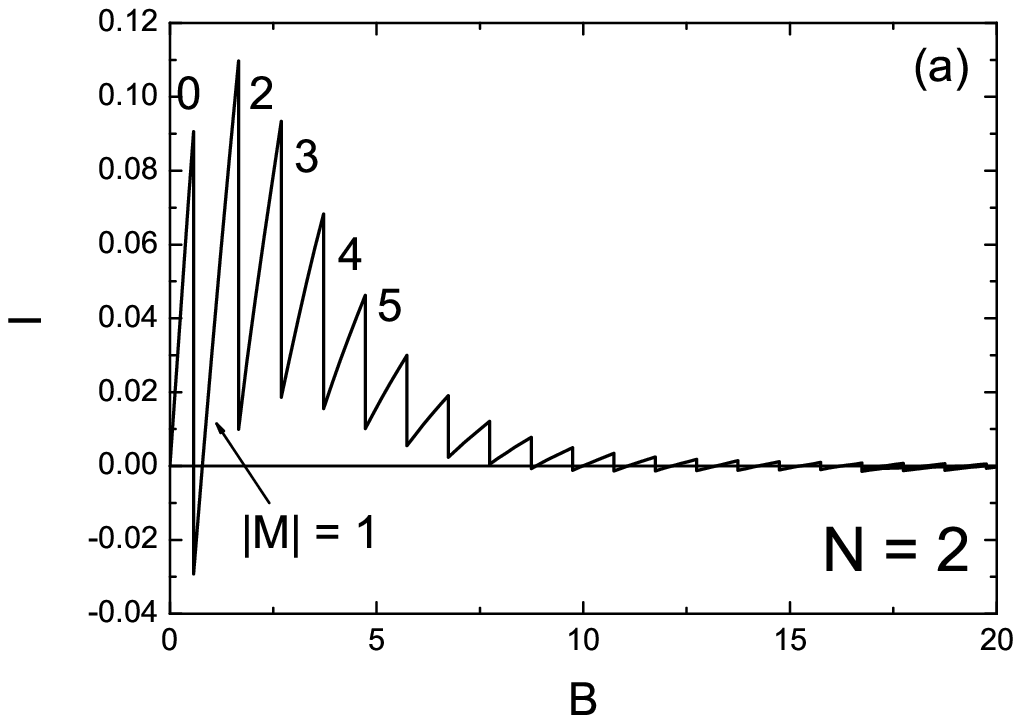}
\includegraphics[width=80mm]{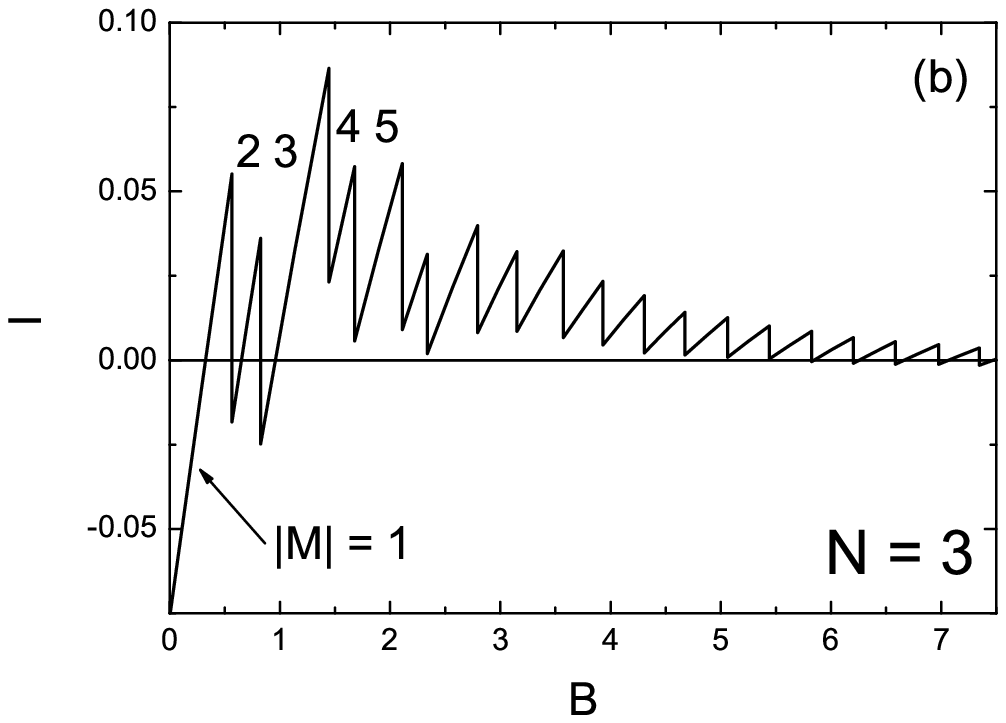}
\includegraphics[width=80mm]{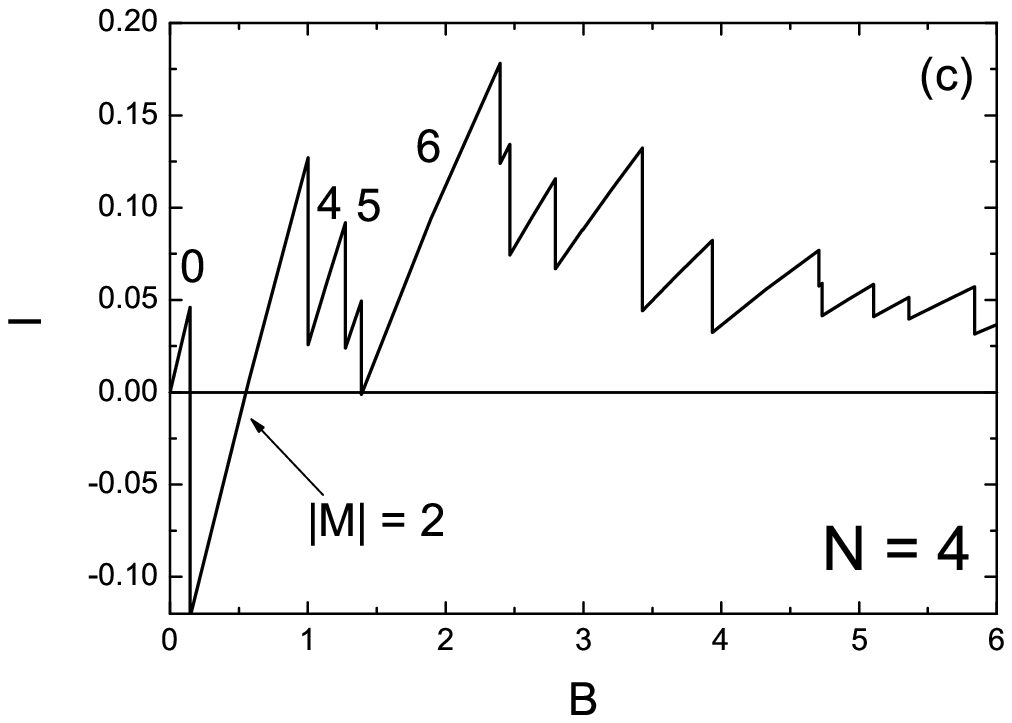}
\caption{\label{fig:saw} The net current crossing the radius of
quantum dots containing two, three and four electrons calculated
by exact diagonalization. The angular momenta up to $|M| = 20$
are included. We observe a saw-tooth-like behavior reminiscent
of persistent currents in quantum rings.}
\end{figure}

Also, it can be noted already from Fig.\ \ref{fig:dens} that the
radial dependence of the current density is nearly antisymmetric
with respect to the classical radius, and thus, the two currents
flowing in the opposite directions nearly cancel each other. That
is, there is almost no global current running along the electron
ring. In order to investigate this phenomenon in more detail we
numerically calculate the magnetic field dependence of the net
current crossing the dot radius
\begin{equation}
  I = \int_{0}^{\infty} j_{\ph}(r)\,dr
\end{equation}
in the ground state. The results are shown in Fig.\ \ref{fig:saw}.
We consider angular momenta up to $|M| = 20$, and use the value
$\lambda = 4$ for two- and three-electron dots while in the case
of four electrons in the dot we set $\lambda = 2$.

We see that these dependences display a saw-tooth-like behavior.
The net current $I$ increases continuously with the magnetic field
in each ground state of a given angular momentum and drops
abruptly at each angular momentum transition reminding persistent
currents in a ring.\cite{butt83} These currents are responsible
for the dot magnetization which demonstrates a similar
behavior.\cite{maksym92} Note that in two- and four-electron
quantum dots where the ground state at low magnetic field
has the angular momentum\cite{maarten03} $M = 0$ there is no net
current at zero magnetic field. In contrast, the $B = 0$ ground
state angular momentum in three-electron quantum dots\cite{mikh01}
is $|M| = 1$ and we observe a large current at low fields.

For the sake of reference, we denote the absolute magnitude of
the current jumps by $\Delta I$ and the average value of the
current at the
jump by $\bar{I}$. In general, both $\Delta I$ and $\bar{I}$
decrease monotonically towards zero with increasing magnetic
field with a few notable exceptions occurring at very low values
of the magnetic field. That is, the current oscillations tend to
decrease in magnitude with increasing $B$ and the oscillatory
saw-tooth pattern becomes centered around $\bar{I} = 0$. The case
of two electrons is rather regular and thus easy to analyse.
Here, $\bar{I}$ reaches values very close to zero already at
angular momenta around $|M| \approx 10$ while the behavior of
three- and four-electron dots is more complicated. There are
conspicuous irregularities associated with more stable states
at the magic values of angular momentum, and moreover, the
oscillations do not center around $\bar{I} = 0$ up to higher
magnetic fields. For three-electron dots $\bar{I}$ approaches
zero only at highest values $|M| \approx 20$ while in
four-electron dot the current oscillations do not center
around zero in the considered parameter range at all.

We find that the dependence of $\bar{I}$ on the magnetic field
is approximately exponential which indicates its essentially
quantum-mechanical nature. Therefore, this aspect of current
oscillations can not be captured by our simple quasi-classical
model whose prediction is $\bar{I} \equiv 0$. But in contrast,
the behavior of $\Delta I$ can be analysed and understood from
a classical point of view.

For this purpose, it is sufficient to approximate the angle
$\chi$ as
\begin{equation}
  \chi \approx \frac{1}{N} \sum_{n=1}^N \ph_n,
\end{equation}
and the density of the electron current along the ring can be
calculated as $N$ times the average of the one-electron velocity
operator. Its sole azimuthal component reads
\begin{equation}\label{currdens}
  j_{\ph}(\br) = \left(\frac{M}{r}+\frac{1}{2}BNr\right)\rho(\br),
\end{equation}
and the total current can be estimated as
\begin{equation}\label{totcurr}
\begin{split}
  I &= \int_0^{\infty}dr\left(\frac{M}{r}+\frac{1}{2}BNr\right)\rho(\br)
  \Big/\int d^2r \rho(\br) \\
  &\approx \left(\frac{M}{a}+\frac{1}{2}BNa\right)
  \int_0^{\infty}dr\rho(\br)\Big/\int d^2r\rho(\br) \\
  &\approx \frac{1}{2\pi a^2}
  \left(\frac{M}{a}+\frac{1}{2}BNa\right).
\end{split}
\end{equation}
Thus, the current is proportional to the largest factor in the
potential (\ref{potzero}) term which was assumed to be zero when
defining the approximate ring radius $a_0$. It means that for the
calculation of the persistent current the radius of the ring has
to be defined with a greater precision. So, let us equate
the derivative of Eq.~(\ref{potzero}) to zero
\begin{equation}\label{derzero}
  \frac{d}{da}V_0 = -\frac{1}{Na^3}\left(M^2-\frac{B^2N^2a^4}{4}\right)
  + N\left(a - \frac{N\lambda f_N}{2a^2}\right) = 0.
\end{equation}
Using this expression together with Eq.~(\ref{mombrel}) we present
the current (\ref{totcurr}) as
\begin{equation}\label{curr}
\begin{split}
  I &= \frac{N^2a}{2\pi}\left(M-\frac{1}{2}BNa^2\right)^{-1}
  \left(a - \frac{N\lambda f_N}{2a^2}\right) \\
  &= - \frac{N}{2\pi B}\left\{1 - \frac{\lambda f_N}{2}
  \left(\frac{NB}{2|M|}\right)^{3/2}\right\}.
\end{split}
\end{equation}
We observe that the magnetic field can not be compensated exactly
by a discrete value of the orbital momentum. Thus, we define the
critical values of the magnetic field $B_M$ that correspond to the
exact compensation as
\begin{equation}\label{bcrit}
  B_M = \frac{2|M|}{Na_0^2} = \frac{2|M|}{N}(\lambda f_N/2)^{-2/3}.
\end{equation}
Then, introducing the magnetic field deviation $B = B_M+\Delta B$
we rewrite the expression for the current (\ref{curr}) as
\begin{equation}\label{currf}
  I = -\frac{N}{2\pi B_M}\left\{1 - \left(\frac{B_M+\Delta B}
  {B_M}\right)^{3/2}\right\} \approx
  \frac{3N\Delta B}{4\pi B_M^2}.
\end{equation}
Finally, estimating the value of the current jump at the angular
momentum transitions we substitute
$\Delta B^{(\mathrm{max})}=2/Na_0^2$
into Eq.~(\ref{currf}) and find
\begin{equation}\label{currj}
  \Delta I = \frac{A_N}{\lambda^{2/3} B_M^2}, \quad
  A_N = \frac{3}{2\pi(f_N/2)^{2/3}}.
\end{equation}

\begin{figure}
\includegraphics[width=80mm]{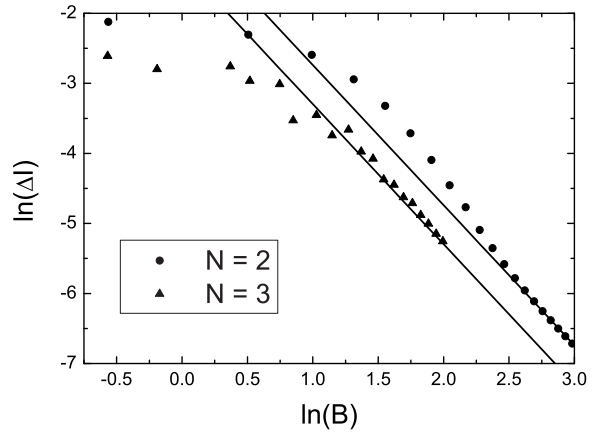}
\caption{\label{fig:power} Double-logarithm plot of the current
jumps $\Delta I$ versus the dimensionless magnetic field $B$.
The symbols (full lines) denote the numerical exact diagonalization
(quasi-classical) results.}
\end{figure}

Thus, our theory predicts a power-law dependence of the current
jumps on the magnetic field $\Delta I \sim B^{-2}$. In order to
test this prediction against the results of the numerical
calculations, in Fig.\ \ref{fig:power} we plot $\Delta I$ as a
function of $B$ for quantum dots containing two and three
electrons. The symbols depict the numerical results from our
exact diagonalizations while the straight lines describe the
quasi-classical limit. One sees, that at high magnetic fields
the numerical results for two- and three-electron dots indeed
cross into a linear regime in a full agreement with
Eq.~(\ref{currj}).

The distribution of currents in quantum dots shown in
Fig.\ \ref{fig:dens} and the appearance of saw-tooth-like
oscillations can be explained on the basis of the following
simple considerations. The physical current consists of two
components, the so-called paramagnetic current proportional
to the gradient of the phase of the wave function and the
contribution due to the vector-potential. The former component
resembles the irrotational flow of the Euler liquid known in
hydrodynamics. This current behaves as $j_{\ph} \propto m/r$
where $m$ is the electron orbital momentum, and therefore, its
vorticity (curl) is zero everywhere except at the origin. On
the other hand, the vector-potential part introduces a
rigid-body-like rotation with $j_{\ph} \propto r$, and
the vorticity of this component is the same at every point.
As any physical system tries to minimize its energy and tends
to have as small currents as possible, the two components
are required to cancel each other. However, due to the very
different $r$-dependences such cancelation is possible only at
the classical radius of the quantum dot.

The net integrated
current crossing the radius of the quantum dot is also, in
general, non-zero. According to Eq.\ (\ref{mombrel}), an exact
cancelation of the net current can be realized only for a
specific value (close to the classical value) of the electron
ring radius. The Coulomb repulsion between the electrons
precludes the adjustment of the radius thereby causing the
appearance of the uncompensated persistent current.
This many-electron effect is essentially different from the
single electron models\cite{tan99} where persistent currents
appear due to fixation of the electron ring radius by the
confinement potential or system boundaries.

\section{Local currents}

In order to obtain a better understanding of the competing
currents flowing in the opposite directions on the inner and
outer edges of the electron ring, we will now consider the
density-current correlation functions which could also be
called conditional currents. In close similarity to the more
familiar density-density correlators (conditional densities),
these functions are obtained by pinning one of the electrons
at a certain point and inspecting the distribution of currents
created by the other electrons in the dot. In our calculations,
we place the pinned electron at the distance equal to the
classical radius $a_0$ from the dot center. The remaining
electrons will tend to localize close to their
crystallization points, distributed equidistantly along
the circumference of the ring. In the quasi-classical wave
function (\ref{wff}) the electrons are distinguishable and
the motion of each of them is restricted to the vicinity of
its own crystallization point. Therefore, when evaluating
the quasi-classical correlation function close to these
points we may dispose with the summation over the electrons
and take into account only the nearest one.

When considering the correlation functions, the rotating frame
coordinates $(x_n, y_n)$ are not the most convenient ones due to
the presence of the constraint (\ref{cond1}). Thus, having a
pinned electron does not imply fixed values of its coordinates.
In order to get rid of this nuisance we switch to a new set of
coordinates $(\xi_n, \eta_n)$ which measure, respectively, the
radial and angular deviations of the electrons from their
crystallization positions in the laboratory frame. In the
considered complex representation (\ref{transf}) both
coordinate sets are related as
\begin{eqnarray}\label{coordrel}
&&  \exp(i[\alpha_n+\chi])(a + x_n + iy_n) \nonumber \\
&& = \exp(i\alpha_n)(a + \xi_n + i\eta_n).
\end{eqnarray}
Separating the real and imaginary parts we are immediately led
to the following two equations of the coordinate transformation
\begin{subequations}\label{trans}
\begin{eqnarray}
\label{trans1}
  x_n + a &=& (\xi_n + a)\cos\chi + \eta_n\sin\chi, \\
\label{trans2}
  y_n &=& \eta_n\cos\chi - (\xi_n + a)\sin\chi.
\end{eqnarray}
\end{subequations}
Performing the summation over all electrons in Eq.~(\ref{trans2})
and fulfilling the constraint (\ref{cond1}) we define the rotation
angle $\chi$
\begin{equation}
\label{tanchi}
  \tan\chi = \frac{\sum_{n=1}^N \eta_n}{Na_0+\sum_{n=1}^N \xi_n}.
\end{equation}
Now, inserting this expression into Eqs.~(\ref{trans}) we solve
them for ($x_n, y_n$). Further inserting the solutions into
Eq.~(\ref{wff}) we obtain the quasi-classical wave function in the
laboratory frame. Taking into account the fact that the
quasi-classical wave function was derived in the quadratic
approximation and correspondingly retaining only the necessary terms
\begin{subequations}\label{terms}
\begin{eqnarray}
  \label{termsa}
  \chi &=& \frac{\sum_{n=1}^N \eta_n}{Na_0}
	\left[1 - \frac{\sum_{n=1}^N \xi_n}{Na_0}\right],\\
  \label{termsb}
  x_n &=& \xi_n, \\
  \label{termsc}
  y_n &=& \eta_n - \frac{1}{N}\sum_{n=1}^N\eta_n,
\end{eqnarray}
\end{subequations}
we obtain the wave function in the laboratory frame as
\begin{equation}
\label{ncfun}
  \Psi = e^{iM\chi} e^{-B\tilde{K}/4}.
\end{equation}
Here, the function $\tilde{K}$ is defined by the same expression
as Eq.~(\ref{wff2}) with the variables ($x_n,y_n$) replaced by
the laboratory-frame coordinates ($\xi_n,\eta_n$), and the angle
$\chi$ is now given by Eq.\ (\ref{termsa}).

We begin by investigating the shape of the electron density
lumps formed in the vicinity of their crystallization points
as these results will turn out handy when discussing the
currents. Straightforward integration of the $N$-particle density
\begin{equation}\label{ndens}
  \rho_N = |\Psi|^2 = e^{-B\tilde{K}/2}
\end{equation}
over the coordinates of $N-2$ electrons with the $N^{\mathrm{th}}$
electron fixed (i.~e., $\xi_N=\eta_N=0$) and the first electron
coordinates renamed into $\xi,\eta$, gives
\begin{eqnarray}
\label{denscorr}
  \rho(\xi,\eta) &=& \int d\xi_2 d\eta_2 \cdots d\xi_{N-1}
  d\eta_{N-1} \,\rho_N \nonumber\\
  &=& \exp\left[-\frac{B}{4}\left(\frac{2N-1}{N-1}\xi^2 +
  \eta^2\right)\right]
\end{eqnarray}
That is, the electron lump has the form of a Gaussian elongated
in the azimuthal direction. The contour lines connecting the
constant-density points are ellipses with the ratio of semiaxes
\begin{equation}
\label{ratio1}
  p_N = \sqrt{\frac{2N-1}{N-1}} = \sqrt{3}, \sqrt{5/2}, \sqrt{7/3}
  \quad  \textrm{for } N = 2,3,4.
\end{equation}

We calculate the current-density correlation function in two steps.
First, we consider the total current in the $N$-particle space and
write out its component due to the first electron
\begin{equation}\label{wholecurr}
\begin{split}
&  \bj_{1,N} = \{M\nabla_1\chi + {\bf A}\}\rho_N \\
&  =  \Bigg\{\left[-\frac{M\sum_{n=1}^N\eta_n}{(Na_0)^2}
  - \frac{B}{2}\eta_1\right]\be_{\xi} \\
&  + \left[\frac{M}{Na_0}-\frac{M\sum_{n=1}^N\xi_n}{(Na_0)^2}
  + \frac{B}{2}(a_0+\xi_1)\right]\be_{\eta}\Bigg\}\rho_N \\
& = \bj_{1,N}^{(p)} + \bj_{1,N}^{(l)}.
\end{split}
\end{equation}
Here, the symbols $\be_{\xi}$ and $\be_{\eta}$ denote the unit
coordinate vectors. The term independent of $\xi$ and $\eta$
\begin{equation}\label{persistn}
  \bj_{1,N}^{(p)} = \left(\frac{M}{Na_0}+ \frac{Ba_0}{2}\right)
  \be_{\eta}\rho_N
\end{equation}
leads to the persistent current which was already discussed above.
Due to the integer-valued quantization of $M$ this current vanishes
only at the special values of the magnetic field strength $B_M$
given by Eq.~(\ref{bcrit}). Using Eq.~(\ref{mombrel}) the term
linear in $\xi$ and $\eta$ describing the local currents can be
rewritten as
\begin{equation}\label{localcurr}
\begin{split}
& \bj_{1,N}^{(l)} = \frac{B}{2}\Bigg\{
  -\left[\eta_1 - \frac{1}{N}\sum_{n=1}^N\eta_n\right]\be_{\xi} \\
& +\left[\xi_1 + \frac{1}{N}\sum_{n=1}^N\xi_n\right]
  \be_{\eta}\Bigg\}\rho_N
  = -\frac{1}{2}[{\bf e}_z\times\nabla_1]\rho_N,
\end{split}
\end{equation}
where ${\bf e}_z$ is the unit vector along the direction of the
magnetic field.
Note that this relation has the same form as that previously
obtained for single-electron densities and currents.\cite{note}

Now we are ready to perform the second step and calculate the
correlator of the local currents by integrating the obtained
expression (\ref{localcurr}) over the coordinates of $N-2$ electrons
and pinning the $N^{\mathrm{th}}$ electron as it was done in
obtaining Eq.~(\ref{denscorr}). This procedure amounts to the
replacement of the $N$-particle density $\rho_N$ by the density
correlator (\ref{denscorr}) because the integration does not
involve the coordinates of the remaining first electron. Thus,
dropping the indices and denoting the first electron coordinates
by ($\xi,\eta$) we arrive at the final expression for the
current-density correlation function
\begin{equation}\label{currdens2}
  \bj(\xi,\eta) = -\frac{1}{2}[{\bf e}_z\times\nabla]\rho(\xi,\eta).
\end{equation}

The obtained simple expression leads to important consequences. The
current lines are perpendicular to the gradient of the density, and
\begin{equation}\label{div}
  \mathrm{div}\,\bj(\xi,\eta) =
  -\frac{1}{2}\nabla[{\bf e}_z\times\nabla]\rho(\xi,\eta) = 0.
\end{equation}
That is, in the considered quasi-classical approximation the local
currents circulating around the localized electrons are conserved.
Therefore, they are physically well defined even though there is no
general conservation theorem for the conditional currents
(see Appendix \ref{appenda}).

This circulatory motion of electrons is similar to the cyclotron
rotation present in any single-electron model (see, e.~g., Ref.\
\onlinecite{lent91}). However, due to the electron correlation and
according to Eq.~(\ref{currdens2}) the electrons rotate not along
Larmor circles but along elliptic density contour lines.
The standard quantity describing
such rotational motion is the vorticity (curl) of the current field.
In our case it reads
\begin{equation}\label{vorticity}
\begin{split}
  \mathrm{curl}\,\bj(\xi,\eta) &=
  -\frac{1}{2}[\nabla\times[{\bf e}_z\times\nabla]]\rho(\xi,\eta)
  = -\frac{1}{2}\nabla^2\rho(\xi,\eta) \\
  &= \frac{B}{4}\{p_N^2+1 - B(p_N^4\xi^2+\eta^2)\}\rho(\xi,\eta).
\end{split}
\end{equation}
It is worth pointing out that the vorticity is positive (i.~e., it
has the same sign as the vorticity of the current component due to
the vector potential) at the electron crystallization points, and
becomes negative at a certain distance away from them. The
constant-vorticity contours are more elongated in the azimuthal
direction than the density contour lines, in particular, the contour
corresponding to the zero vorticity has an elliptic shape with the
ratio of semiaxes equal to $p_n^2$ rather than $p_n$.

\begin{figure}
\includegraphics[width=70mm]{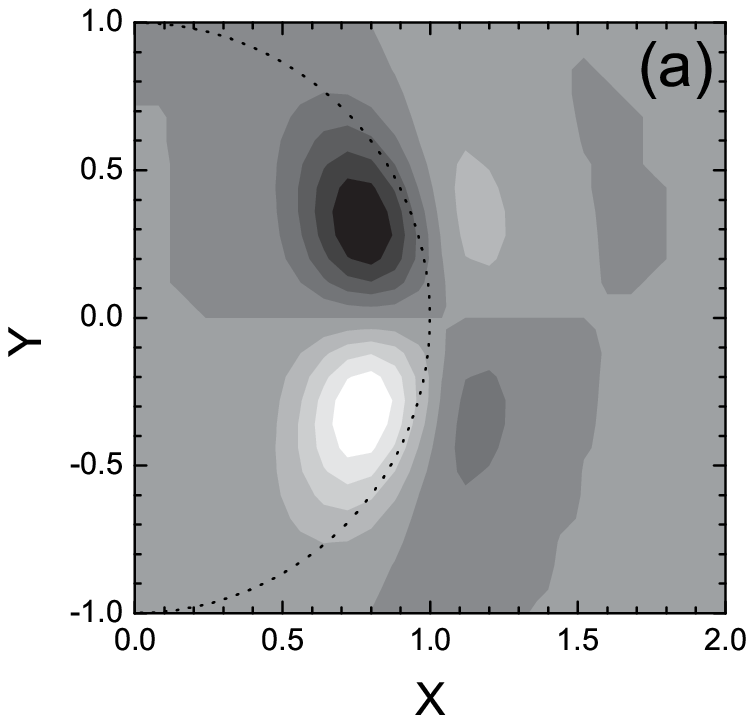}
\includegraphics[width=70mm]{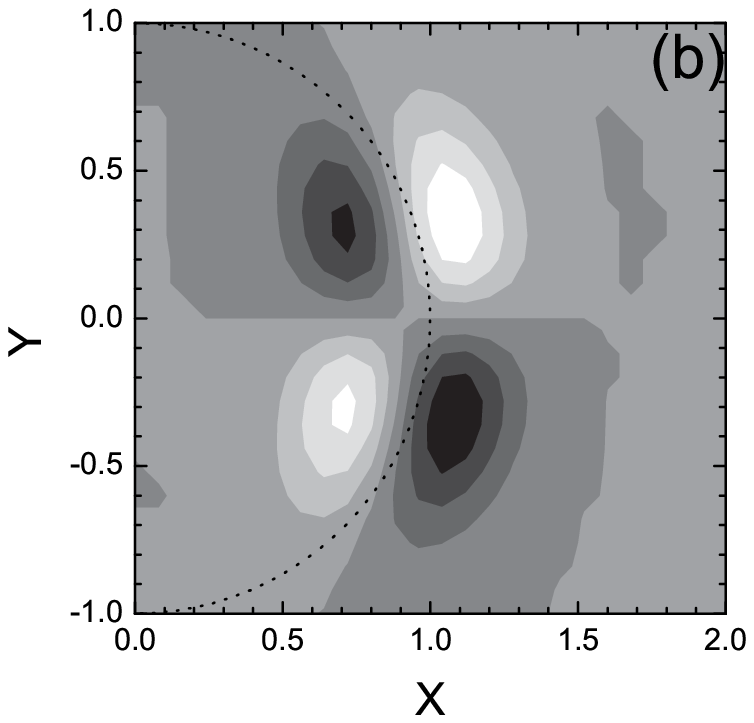}
\includegraphics[width=70mm]{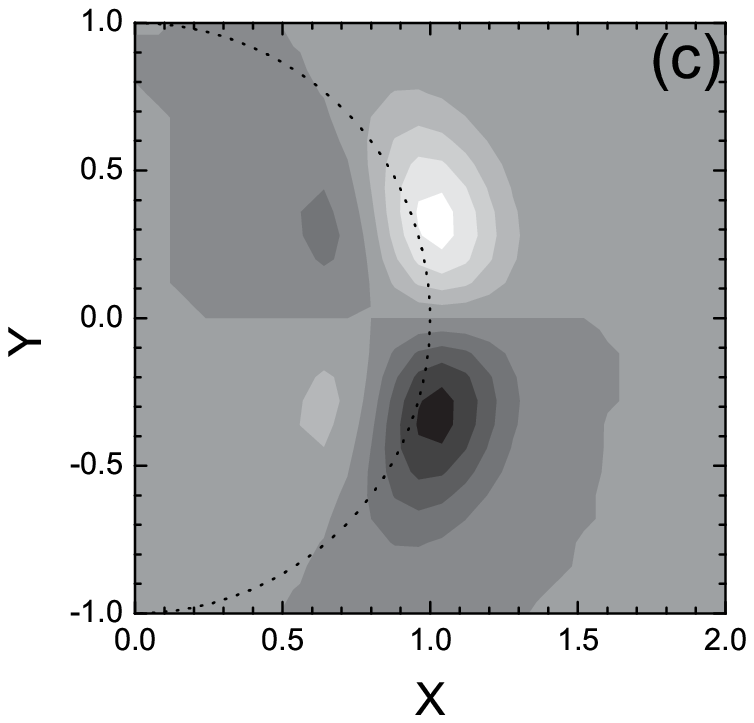}
\caption{\label{fig:div2} The divergence of the density-current
correlator in a two-electron quantum dot. The angular momentum is
$M = -18$. Panel (a) corresponds to the the lower end
(i.~e., $B = 17.75$) of the magnetic field interval where this
state is the ground state, panel (b) is plotted for the medium
value ($B = 18.25$), and panel (c) is obtained at the higher end
($B = 18.75$) of this interval. Uncompensated global currents
are visible. Dark (light) areas correspond to generation
(extinction) of the current. The dotted curve denotes the classical
radius of the quantum dot.}
\end{figure}

We illustrate the above conclusions with numerical results.
The divergence of the density-current correlation function
calculated for a two-electron dot is plotted in Fig.\ \ref{fig:div2}.
We choose the value $\lambda = 4$ so that the classical quantum
dot radius (indicated by a dotted line in the plots) is $a_0 = 1$,
and place the pinned electron
at $(-1,0)$. Dark (light) areas correspond to the positive
(negative) divergence, i.~e.\ generation (extinction) of
the current. The three panels of Fig.\ \ref{fig:div2} show the
same ground state with $|M| = 18$, however, at slightly different
magnetic fields. The plot in panel (a) is obtained for the lowest
possible magnetic field at which this state is the ground state
($B = 17.75$), panel (b) corresponds to the middle of this
interval of magnetic fields ($B = 18.25$), and panel (c) is
obtained at the higher limit of this range ($B = 18.75$).
In all three plots non-conserved currents are visible. Panels (a)
and (c) show a dipole-like structure corresponding to the azimuthal
persistent current (\ref{persistn}). At low magnetic fields [panel (a)]
there is a net current running in the clockwise direction and at
high magnetic fields we observe a net counter-clockwise current,
in accordance with Eq.~(\ref{persistn}) and the saw-tooth-like
behavior of Fig.\ \ref{fig:saw}(a). The data in panel (b) is obtained
at the middle of the allowed interval of the magnetic fields. Here,
$B = B_M$ and there is no global current. However, the divergence
assumes a quadrupole-like checkerboard pattern still indicating
the presence of a certain current non-conservation. This is a small
higher-order effect not captured by the above quasi-classical
treatment.

\begin{figure}
\includegraphics[width=70mm]{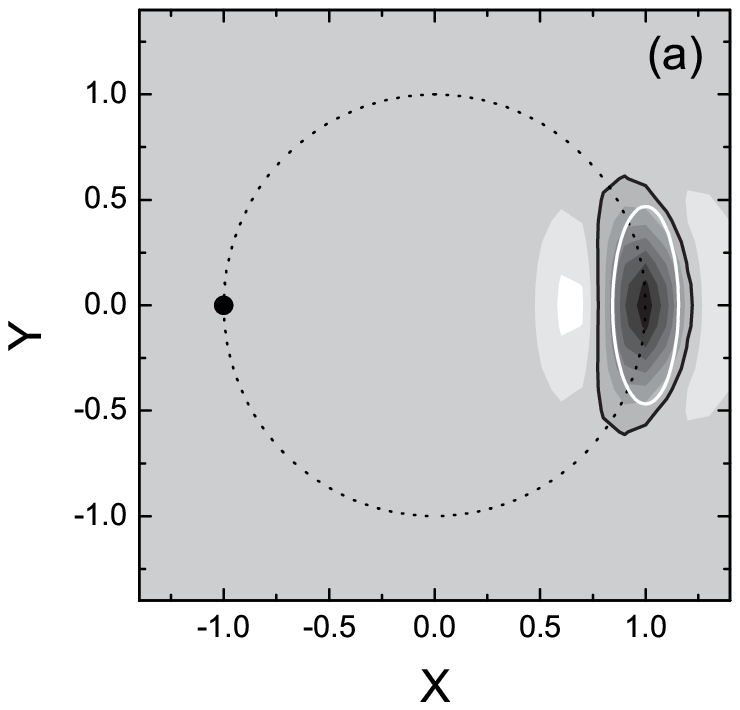}
\includegraphics[width=70mm]{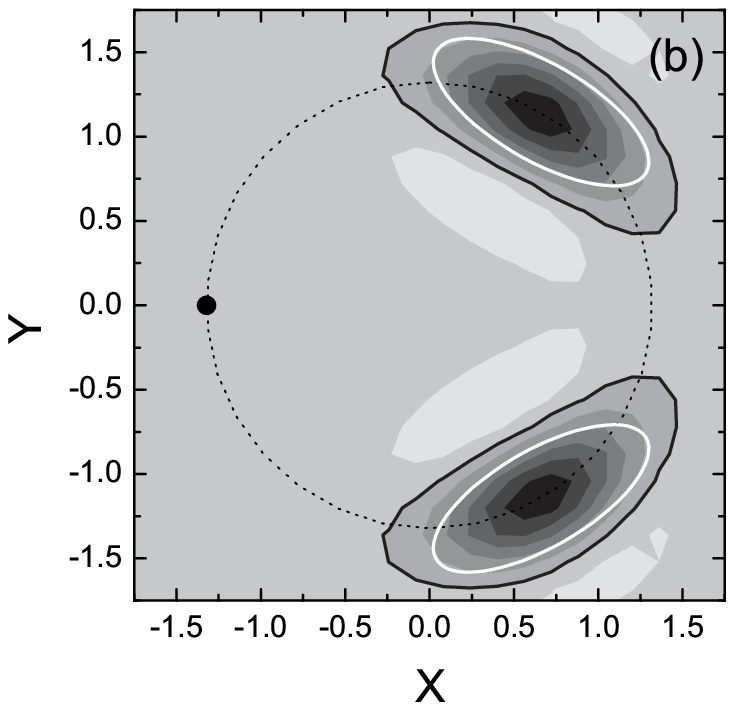}
\includegraphics[width=70mm]{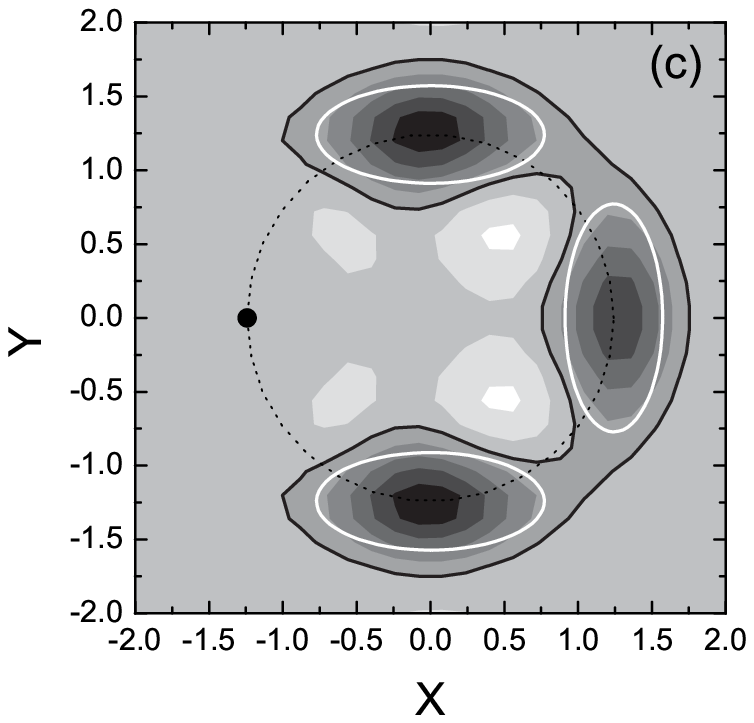}
\caption{\label{fig:curl} The curl of the density-current
correlation function for quantum dots containing two, three
and four electrons. The ring of the classical radius is
marked by the dotted line, and one electron is pinned at
the intersection of this ring with the negative part of the
$x$-axis (indicated by the dot). The ground state of $M = -18$
is shown. Dark (light) areas correspond to positive (negative)
vorticity, and the full black line denotes the separating
zero-vorticity contour. The quasi-classical prediction for
this contour is marked by the white line.}
\end{figure}

Fig.~\ref{fig:curl}(a) shows the curl (vorticity) of a two-electron
quantum dot. We find that the absolute values of the curl are $3$
orders of magnitude larger than the divergence, so the current
non-conservation is small indeed and can be safely neglected. Here,
only one plot corresponding to the middle of the allowed energy
range [as in Fig.~\ref{fig:div2}(b)] is shown since these plots
are rather insensitive to small variations of the magnetic
field. The first electron is pinned at the point $(-1,0)$ on
the classical radius (the dotted line), and the second electron
is crystallized in the vicinity of the opposite point $(1,0)$
and performs a cyclotron-resonance-like motion. Dark (light)
areas correspond to the positive (negative) vorticity, and the
full black line separates the areas corresponding to different
signs of the curl. According to the theory predictions, this
line is an elongated ellipsis and is marked by a white line.
We see that the numerical results indicate a certain spreading
and deformation of this area as the quasi-classical regime is
not truly reached.

For three and four electrons we also obtain a dipole-like structure
in the divergence plots since at the considered magnetic fields
and angular momenta the global rotation of the electron molecule
is not fully stopped, i.~e.\ the oscillations of the current
[see Figs.\ \ref{fig:saw}(b) and \ref{fig:saw}(c)] are not centered
around zero value. In these cases, the maximum absolute values of the
divergence are around $25$ times smaller than the corresponding
magnitudes of the curl. The plots of the curl are shown in
Figs.\ \ref{fig:curl}(b) and \ref{fig:curl}(c) for three and four
electrons, respectively. We again consider the states with the
angular momentum $M = -18$, and set the magnetic field strength
to the midpoint value of the field range corresponding to the
considered ground state. One electron is pinned on the negative
part of the $x$-axis at the distance equal to the classical radius
(marked by the dotted line) from the center. For three electrons
with $\lambda = 4$ this radius is $a_0 = 1.322$, and for the
four-electron dot with $\lambda = 2$ we have $a_0 = 1.242$. The
plots in Figs.\ \ref{fig:curl}(b) and \ref{fig:curl}(c)
again show the remaining electrons localizing close to their
crystallization positions and performing the cyclotron-like
motion along trajectories elongated in the azimuthal direction.
The numerically calculated zero-vorticity contours are expanded
and distorted with respect to the ones predicted by the
quasi-classical approximation. In the case of four electrons
in a dot the regions of positive vorticity even overlap.

\section{Conclusion}

In conclusion, we developed a quasi-classical theory based on a
transformation to a rotating frame that is capable of describing
the currents in quantum dots at high magnetic fields. The results
show that due to the competition between the paramagnetic and
vector-potential components, there arise global persistent
currents running along the electron ring. At the same time,
each electron may be visualized as performing individual
cyclotron-like motion along elliptic trajectories elongated
in the azimuthal direction. This motion is well described by
the density-current correlation functions. We introduced these
functions as a refinement of the suggestion to demonstrate the
Wigner crystallization by calculating ordinary currents in the
broken-symmetry situation, and showed that in the quasi-classical
limit they are conserved, and thus, physically well defined.

\acknowledgments

This work is supported by the Belgian Interuniversity Attraction
Poles (IUAP), the Flemish Science Foundation (FWO-Vl), VIS (BOF)
and the Flemish Concerted Action (GOA) programmes.
E.A.\ is supported by the EU Marie Curie programme under contract
number HPMF-CT-2001-01195.

\appendix

\section{Two electrons on a 1D ring}
\label{appenda}

In this section, we present a simple solvable model that, by means
of a geometrical interpretation, illustrates the problems arising
with the single-electron functions and correlators when discussing
Wigner crystallization.

We consider two electrons moving on a 1D ring of radius $r=1$ in a
perpendicular magnetic field whose behavior is described by the
following Hamiltonian
\begin{equation}\label{ham}
  H = -\frac{1}{2}\left\{\left(\frac{\p}{\p\ph_1}+i\gamma\right)^2
  + \left(\frac{\p}{\p\ph_2}+i\gamma\right)^2\right\}
  + \lambda\cos(\ph_1-\ph_2).
\end{equation}
Here, the positions of the electrons on the ring are given by the
angles $\ph_{1,2}$, the symbol $\gamma = eBr^2/2c\hbar$ stands for the
dimensionless vector potential, and the electron-electron interaction
is modeled by the cosine function which ensures the Wigner
crystallization of the electrons on the opposite ends of the diameter
in the strong interaction ($\lambda\to\infty$) limit.

\begin{figure}
\includegraphics[width=70mm]{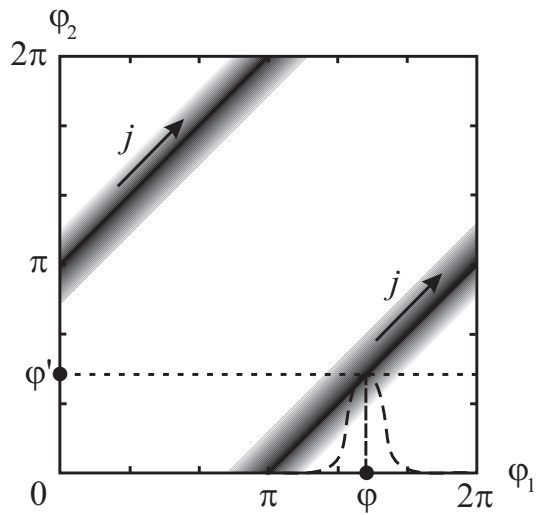}
\caption{\label{fig:tworing} Many-electron density.}
\end{figure}

After the transformation
\begin{equation}\label{newcoord}
  \Phi = \frac{1}{2}(\ph_1+\ph_2), \quad \ph = \ph_1 - \ph_2 + \pi
\end{equation}
the variables separate, and we rewrite the Hamiltonian (\ref{ham})
in the $\lambda\to\infty$ limit as
\begin{equation}\label{apham0}
  H = -\frac{1}{4}\left(\frac{\p}{\p\Phi} + 2i\gamma \right)^2
  - \frac{\p^2}{\p\ph^2} + \frac{\lambda}{2}\ph^2.
\end{equation}
whose ground-state eigenfunction is
\begin{equation}\label{wf}
  \Psi(\ph_1,\ph_2) = e^{iM(\ph_1+\ph_2)/2}\,
  e^{-(\ph_1-\ph_2+\pi)^2\sqrt{\lambda/8}}.
\end{equation}
Here, the symbol $M$ stands for the total angular momentum.
In the ground state, its value equals to the integer closest
to $-2\gamma$. The corresponding many-electron density
\begin{equation}\label{density}
  \rho(\ph_1,\ph_2) = e^{-(\ph_1-\ph_2+\pi)^2\sqrt{\lambda/2}},
\end{equation}
is shown in Fig.\ \ref{fig:tworing} by the dark stripes in the
2D many-electron space.
In this space, the current has two components related to both
electrons. Its first component is given by the expectation value
of the first electron velocity operator $(-i\p/\p\ph_1+\gamma)$
\begin{equation}\label{fcurr}
  j_1 = (M/2 + \gamma)\rho(\ph_1,\ph_2).
\end{equation}
The same result is obtained for the second component of the total
current $j_2$. Thus, the current flows along the dark stripes as
indicated by arrows in Fig.\ \ref{fig:tworing}.

Eqs.~(\ref{density}), (\ref{fcurr}) and Fig.\ \ref{fig:tworing}
contain all available information about the system. However, usually
it is too difficult to consider many-electron spaces, and one resorts
to simpler single-electron functions such as the density and the
current obtained by integrating of Eqs.~(\ref{density}) and (\ref{fcurr})
over the coordinates of the second electron
\begin{subequations}\label{sfunc}
\begin{eqnarray}
  \rho(\ph) &=& \int_0^{2\pi}d\ph'\rho(\ph,\ph') = \mathrm{const}, \\
  j(\ph) &=& (M/2+\gamma)\int_0^{2\pi}\!\!d\ph'\rho(\ph,\ph') \nonumber\\
  &\varpropto& M/2+\gamma.
\end{eqnarray}
\end{subequations}
The current expression gives the persistent current flowing along
the ring. From the geometrical point of view, the above expressions
are projections of the 2D many-electron density onto the abscissa
axis. Naturally, they lead to a homogeneous single-electron density
which provides no information about the Wigner crystal.
Thus, one has to consider more sophisticated correlation functions.
In our case, the density-density correlation function is given
by Eq.~(\ref{density}) and the density-current correlation function
by Eq.~(\ref{fcurr}). In both cases, they are treated as functions
of the first electron coordinate with the second electron coordinate
fixed at the point $\ph_2=\ph'$. Geometrically this actually means
taking the cross-section of the plot shown in Fig.\ \ref{fig:tworing}
along the dotted line $\ph'=\mathrm{const}$. The density-density
correlation function is shown by a dashed line, and we see a lump
at the point $\ph_1=\ph=\ph'+\pi$ which serves as a good indication
of the formation of a Wigner crystal.

The current-density correlation function is not so fortunate
because of the non-zero divergence
\begin{equation}\label{apdiv}
  \frac{d}{d\ph}j_1(\ph,\ph') = -\sqrt{\lambda/2}(M/2 + \gamma)
  (\ph-\ph'+\pi)\rho(\ph,\ph') \ne 0,
\end{equation}
indicating the absence of the charge conservation. It is evident
from Fig.\ \ref{fig:tworing} that the global current does not flow
along the chosen cross-section, and the currents are escaping into
other dimensions.

\end{document}